\documentclass [10pt, a4paper]{article}
\usepackage{osajnl2} 
\begin{document}

\sloppy

\newcommand{\pthz}[1]{
\mbox{$ #1\;\mathrm{pT}/\sqrt{\mathrm{Hz}}$}}

\title {Dual channel self-oscillating optical magnetometer}

\author{J.\ Belfi  $^{1,}$ \footnote[3]{presently at: Dipartimento di Fisica, 
Universit\`a di Pisa, Largo B.\ Pontecorvo 3, 56127 Pisa, Italy}, 
G.\ Bevilacqua, $^{1}$ V.\ Biancalana, $^{1,*}$ S.\ Cartaleva, $^{2}$ 
Y.\ Dancheva, $^{1}$ K.\ Khanbekyan $^{1}$ and L.\ Moi $^{1}$}
\address{$^1$ CNISM-Unit\`a di Siena, Centro Studi Sistemi Complessi 
and Dipartimento di Fisica - Universit\`a di Siena, via Roma  56,
53100 Siena, Italy} 
\address{$^2$ Institute of electronics, BAS, Boul. Tsarigradsko Shosse 72, 
1784 Sofia, Bulgaria}
\address{$^*$Corresponding author: biancalana@unisi.it}

\begin{abstract}
We report on a two-channel magnetometer based on nonlinear
magneto-optical rotation in a Cs  glass cell with buffer gas. The Cs
atoms are optically pumped and probed by free running diode lasers  
tuned to the D$_2$ line. A wide frequency modulation of the pump laser
is used to produce both synchronous Zeeman optical pumping and
hyperfine repumping. The magnetometer works in an unshielded
environment and spurious signal from distant magnetic sources is
rejected by means of differential measurement. In this regime the
magnetometer simultaneously gives the magnetic field modulus and the
field difference.  

Rejection of the common-mode noise allows for high-resolution magnetometry with 
a sensitivity of  \pthz{2}. This sensitivity, in conjunction with
long-term stability and a large  
bandwidth, makes possible to detect  water proton magnetization and
its free induction  decay in a measurement volume of 5~cm$^3$.   
\end{abstract}


\maketitle

\section {Introduction}  
The potential of atomic magnetometers is well known and was established in the
late 50's and the 60's  \cite{Dem57, Bel57, Blo62, Roc69, Tan69} .

There has been a recent revival of the interest in such devices, specifically in
optical atomic magnetometers, also due to the progress made in diode laser 
technology. This revival has led to the achievement of impressive results 
over the last decade \cite{Ale96, Wyn99, Bud02, Aco06},
highlighting the real applicability of the optical
magnetometry due to its high sensitivity, high accuracy, and high
time and spatial resolution (see also the review \cite{Rom07} and the
references therein for a panorama on this progresses). For each
particular application, a different set of specifications have to be
stressed to their extreme limit, and  atomic 
magnetometers are excellent in terms of trade off flexibility,
when seeking appropriate compromises. Optical
magnetometry based on Faraday rotation has recently been demonstrated to be an
impressively powerful tool even in dense media such as liquids
\cite{Sav06}, where optical rotation directly induced
by the nuclear spin polarization has been observed.

The growing interest in this field is also specifically related to the application 
of optical magnetometers as non-inductive detectors of low-field 
nuclear magnetic resonance (NMR) detection and to the close area of 
its remote detection \cite{Sav05}, as an alternative 
to the superconducting quantum interference device approach. Low-field
NMR imaging with 100~msec and millimetre 
size resolution has been demonstrated \cite{Shoyun06}, 
and the impressive limit of 7~fT/$\sqrt{\mathrm {Hz}}$ has been achieved with a 
NMR detector \cite{Sav07} operating with Potassium in the 100~kHz range. 
A miniaturized (chip-scale) atomic magnetometer has been developed, 
working in spin-exchange relaxation-free regime, which can detect 
the proton field from Hydrogen contained in 1~mm$^3$ water 
prepolarized at 1~T field, with measuring time of 1~sec  \cite{Ledb08}.

In the last few years, our research has been devoted to the construction of
magnetometric setups, considering their possible practical
applications and in-field use \cite{And03}. Our goal is to
build simple but reliable setups with good sensitivity and
long-term stability, operating at room temperature, in an 
unshielded room, and using commercially available laser sources.
The challenge of operating in an unshielded environment has led us to
develop differential setups \cite{Bel_cardio07}. We have
chosen Cs D$_2$ line for operation at room
temperature with commercial lasers. This compromise leads to
disadvantages related to the smaller Land\'e factor and to lower 
efficiency in the pumping process, but the advantage of a single isotope,
high density medium at room (or slightly higher) temperature.

The flexibility of the setup presented here is related 
to the presence of two independent light sources, which makes it
possible to precisely determine the optimal optical detuning 
of the probe laser. This configuration is also suitable for analyzing 
and studying subjects of general and basic research, such as the 
collisional properties of the excited states, the interplay of 
hyperfine and Zeeman pumping etc. However, these subjects 
are beyond the scope of this paper.

Light generally plays a double role in the optical magnetometers
\cite{Sut91}, both creating atomic momentum orientation/alignment 
and detecting its time evolution. This evolution (atomic 
momentum precession around the magnetic field to be measured) 
is as {\it free} as the perturbation produced by the probing light 
is negligible.

Frequency or amplitude modulation of the laser radiation are widely used 
to prepare the state at given instants, and to detect the
consequent optical properties of the medium, which are time-dependent due
to the coherently precessing atomic spins. These
properties can be characterized in terms of absorptive, dispersive and
dichroic behaviour.

The external degrees of freedom of the atoms (translational motion) may 
play a dominant role in degrading the sensitivity of an atomic magnetometer
by reducing of the interaction time of atoms - either bringing them out of
the interaction volume, or by causing them to collide destructively with the cell
walls. Buffer gas \cite{Bev73} and anti-relaxation coatings
have been widely and successfully used for decades now as a solution
to these problems. The coating approach demands the production
of very high quality surface covering and a few groups around the
world \cite{Ale96, Graf05, Guz06} have developed
reliable procedures, with which excellent results have been achieved.
Recently, J.\ M.\ Higbie {\it et al.} \cite{Hig06}  built a robust 
self-oscillating setup based on high quality antirelaxation
paraffin-coated cells excited by two non-overlapping parallel beams. 

Aiming to build a robust self-oscillating magnetometer with a buffer gas cell, 
we developed a setup similar to the one described in \cite{Hig06}, but using 
two laser beams crossing inside the cell, as diffusionally slowed motion
inside the buffer gas would make it impossible to pump and detect atoms in
different locations. In fact, in our experimental conditions 
\emph {(90~Torr of Ne as a buffer gas)} the diffusion 
coefficient of Cs can be estimated (see e.g. \cite{Van89}) as being
about 1~$\mathrm{cm^2/s}$. This leads, within a typical coherence time
(in the range $10\div 100$~ms), to typical atomic displacements of $1
\div 3$~mm - an amount of the same order as the beam size.

The use of buffer gas reduces the efficiency of the Zeeman pumping
process, compared to hyperfine pumping, due to the stirring effect of collisions
between excited alkali atoms and buffer gas atoms. This effect
leads to higher hyperfine pumping with respect to Zeeman pumping
and thus to a significant lack of signal, unless appropriate repump radiation is
used. This is what actually occurs in our setup, where a low-power
wide frequency modulated beam synchronously induces Zeeman optical 
pumping and counteracts the hyperfine optical pumping while an unmodulated
probe beam detects the atomic spin precession. 

The use of two separate beams makes it easy to operate the whole system 
as a self-oscillator, considerably widening the response bandwidth
with respect to the scanned regime \cite{Bel07} and thus also making it an 
attractive tool in fields of application that demand a fast response, such as the
detection of oscillating fields produced by nuclear precession in
low-field NMR experiments.  

In our case the two beams are generated by two independent laser
sources, in contrast with the alternative approach based on splitting
an unmodulated beam and then passively modulating one of the two
beamlets, as done in the recent experiments of P.\ D.\ D.\ Schwindt {\it et al.}
\cite{Schw05} and J.\ M. Higbie {\it et al.} \cite{Hig06}).

As described in the next sections, our magnetometer can operate both with
a single sensor, or with a couple of sensors with a differential
response, making simultaneous absolute and differential
magnetometry possible, with the advantage of significant common-mode
noise rejection. The latter is of a great importance when 
measuring slowly varying magnetic fields in an unshielded environment. 

\section{Experimental Setup}
\begin{figure}[htbp]
 \centerline{\includegraphics[width=10cm]{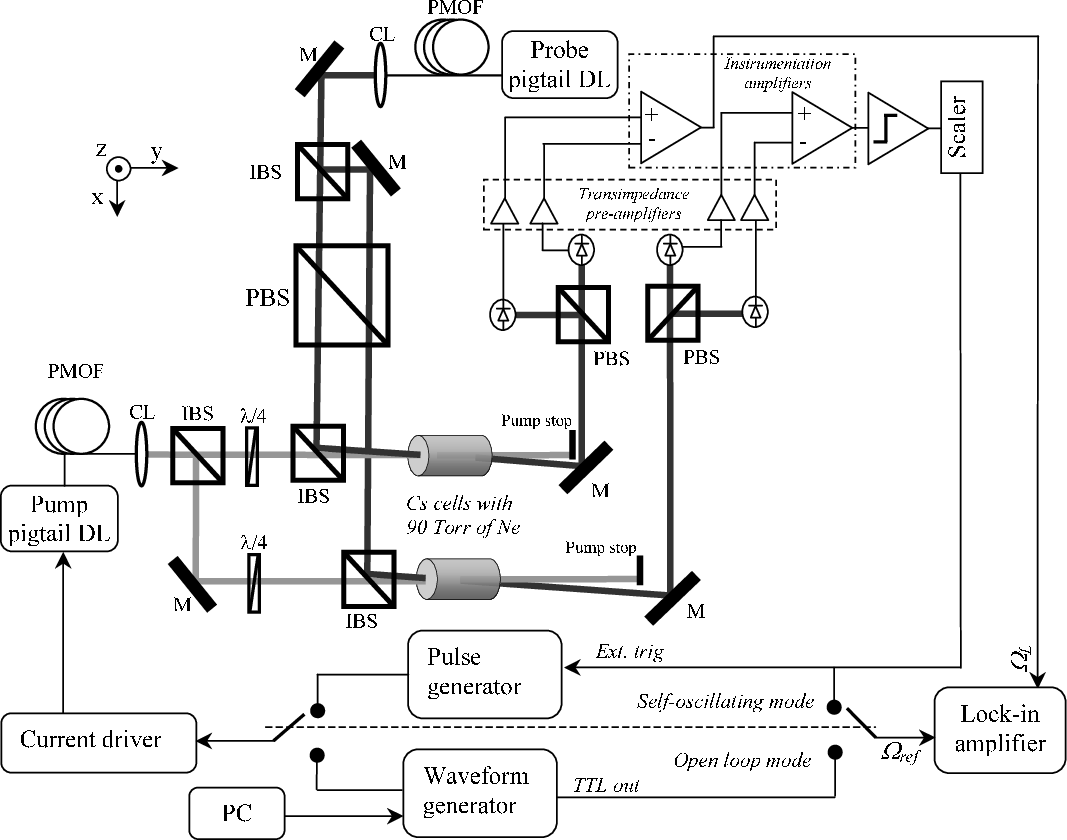}}
    \caption{Schematic of the experimental setup. Abbreviations: PMOF: 
    Polarization Maintaining Optical Fiber,
    PBS: Polarizing Beam Splitter, IBS: intensity beam splitter, M:
    mirror, CL: collimating lens, $\lambda/4$: quarter-wave
    plates. \label{biancalana_f1}}  
\end{figure}

The interaction scheme is suitable for a transverse pumping experiment, 
{\it i.e.} for measurement of a magnetic field (oriented along the $z$
axis) that is perpendicular to the laser beam propagation ($y$
direction) and hence to the laser-induced  orientation of the atomic momentum. 
The experimental apparatus, shown schematically in
Fig.\ref{biancalana_f1}, consists  of two cylindrical Cs cells of 2~cm
in diameter and length, crossed by two laser beams - the 
pump and the probe beam.  
The Cs cells are filled with 90~Torr of Ne and kept at around  $32^\circ$C
using circulating hot water. The probe laser is 1.9~GHz blue-shifted from the
transition starting from $F_g=4$. The two laser beams, tuned to the Cs
$D_2$  line, are supplied by two pigtail diode lasers in free running regime
(UniPhase-SDL laser chip (5411-G1-852), a Faraday isolator
(40~dB) and polarization maintaining fiber assembled by HIV,
Germany). The pump laser frequency is passively stabilized and 3~GHz
red-shifted from the transition starting from $F_g=3$ and is widely
modulated around this frequency position by applying appropriate
pulses at the modulation input of the current driver.

The pump and the probe beams pass through each cell at a small 
angle to each other so that they can be separated at the detection 
stage (around 2~m away), before which the pump beam is blocked. 
Each probe beam, linearly polarized in the
$x$ direction, is analyzed after the cell using a balanced
polarimeter. Each polarimeter is composed of a Wollaston polarizing beam splitter 
(PBS) oriented at 45$^\circ$ with respect to the $z$ and $x$ axes. The
photodiode currents are amplified by two pairs of identical
transimpedance preamplifiers with maximum amplification at around
15~kHz and a bandwidth of about 17~kHz \cite{nota}. The outputs of
each pair of transimpedance amplifiers are fed into an instrumentation 
amplifier (AD620) whose output is monitored with an oscilloscope or acquired.
The signal from the main arm is also used to close the self-oscillating 
loop (to trigger the pulses applied to the pump laser driver) as well as 
for referencing a lock-in amplifier.

\section{Detection principle}
\begin{figure}[htbp]
  \centerline{\includegraphics[width=10cm]{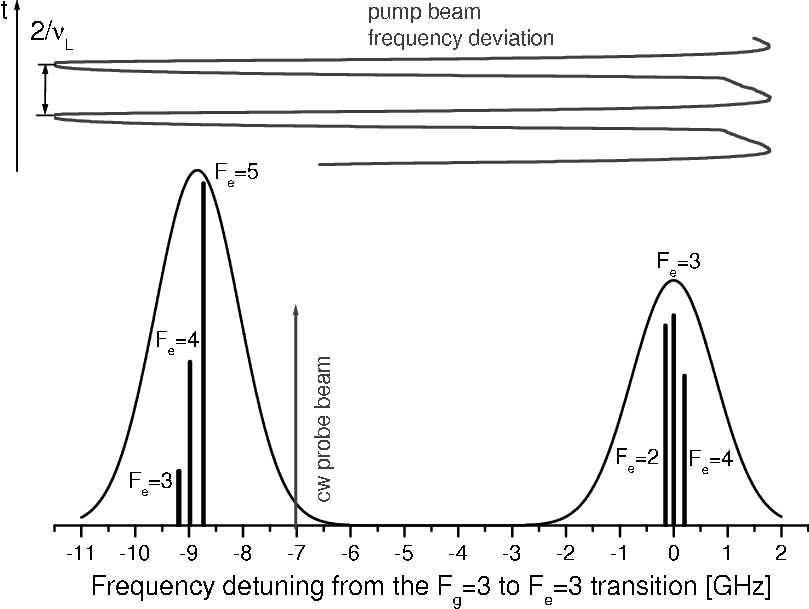}}
    \caption{Instantaneous optical frequency of the pump laser with respect to the 
absorption profile of the Cs D$_2$ line. The distortion of the square wave
that modulates the pump laser frequency is inferred by direct measurement of
the driving current.  \label{biancalana_f2}}
\end{figure}

A detailed representation of the excitation/detection scheme is given
in Fig.\ref{biancalana_f2}. The probe beam is left unmodulated, 
while the pump beam is modulated by applying a pulse sequence at a frequency $\nu$ 
and a duty cycle of $30\%$ to the modulating input of the diode
current driver. Due to the limited bandwidth of the modulation input
(the nominal cutoff frequency is 25~kHz), some distortion of the
pulsed waveform occurs. Fig.\ref{biancalana_f2} shows the absorption
profile of the set of 6 possible transitions for homogeneous
broadening due to high buffer gas pressure, and without considering
the hyperfine optical pumping effect. The center of each transition is
marked, together with its relative oscillator strength. The probe 
beam is blue-detuned with respect to the group of transitions 
starting from $F_g=4$, while the pump beam is broadly modulated, with an
estimated frequency deviation of the order of 14~GHz. The modulation
duty cycle and detuning of the pump laser make it resonant
for short time intervals (pulses) with F$_g=4$ and for
longer time intervals with F$_g=3$. The presence of two laser
sources makes it possible to identify the optimal optical detuning
of each one. The probe laser detuning is important in terms of
resonance contrast and linewidth. The optimal optical detuning 
of the probe laser is found in scanned mode, by looking at the slope
of the resonance curve expressed as A$/\gamma^3$, where the parameters
A and $\gamma$ are the amplitude and the width of the resonance profile, 
respectively, and are determined by means of a best fit procedure with
a Lorentzian curve. The slope decreases asymmetrically around the
maximum at 1.9~GHz and goes down to 80\% at 1.2 and 4.9~GHz
blue-detuning from $F_g=4$ group of lines. The optical frequency of  
the probe laser is passively stabilized at the optimal value, using
the Doppler broadened absorption profile observed in the transmission
of a vacuum Cs cell as a reference. This passive stabilization is  
described in detail in \cite{Bel07} and provides long term drift
correction unless mode-hop occurs, with an uncertainty of 2~MHz for
time scales of 1~sec. The wide frequency modulated pump laser is left unstabilized,
with optical frequency excursions of the order of 200~MHz per
day. An optical frequency drift of 700~MHz from the optimal set point
gives a decrease of the resonance slope of the order of 20\%.

\begin{figure}[htbp]
  \centerline{\includegraphics[width=10cm]{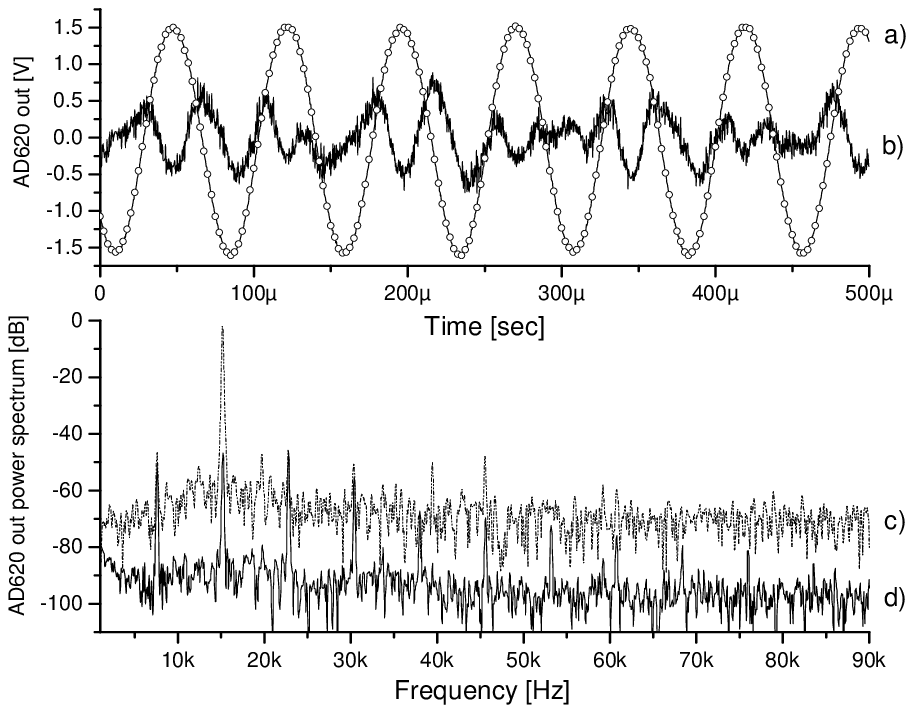}}
    \caption{Signal of one channel at the output of the instrumentation amplifier. 
The upper plot shows the signal {\it vs.} time (trace a) and the residual 
(signal minus its fundamental tone at $\nu_L$) amplified by a factor
10 (trace b). In the lower plot the power spectrum of the signal
(trace c) shows a 60~dB contrast of the peak at  $\nu_L$ with respect
to the background noise.  Trace d shows details of the noise
contributions. This is the power spectrum  of an average of 128
traces, registered with no probe beam. Reduction  of the white noise,
due to averaging, makes it possible to distinguish peaks at the  pump
frequency and at its harmonics. It is worth noting that the harmonics at   
$\nu_L$  is almost 50~dB below the peak at the fundamental tone of
  trace c. \label{biancalana_f3}} 
\end{figure}

The (circularly polarized) pump beam produces Zeeman optical pumping 
by orienting the population  orthogonally to the bias magnetic field. 
After a pump pulse, the atomic spins undergo Larmor precession at
frequency $\nu_L$. The probe beam experiences rotation of the
polarization, whose amount depends on time via the Larmor
precession angle. If the optical pumping rate is modulated at 
frequency $\nu$, then the optical-rotation angle of the probe
polarization will, in general, also oscillate at $\nu$.  The rotation
amplitude is maximum, provided that the pump pulses are synchronous
with the Larmor precession, {\it i.e.} if $\nu$ matches $\nu_L$ or
its sub-harmonics. In the results presented here we force the system
with the first sub-harmonic, {\it i.e.} a pump pulse is applied for
each couple of Larmor cycles. The rotation signal, detected at the
balanced polarimeter and  shown in Fig.\ref{biancalana_f3}, is 
essentially sinusoidal at the  Larmor frequency, which makes
it suitable to trigger the pump pulses in the  self-oscillating mode
without any filtering in the loop \cite{Schw05}. The total harmonic
distortion (ratio between harmonics  and fundamental tone
amplitudes)  is about 2.2\%, including the integer and half-integer
harmonics that originate from stray  light from the pump laser (see
trace d in Fig.\ref{biancalana_f3}). The SINAD, {\it i.e.} the ratio
between the fundamental tone and the residual (noise+harmonics) in
the bandwidth of the transimpedance amplifier, is 31~dB,
corresponding to a ratio of 36 between the rms amplitude of the
fundamental tone and the residual. The fundamental amplitude of
$\pm1.6 {\mbox V_{p-p}}$ corresponds to $\pm 4~\mathrm{mrad}$
optical rotation, so that the measured residual  corresponds to an
uncertainty of 80~$\mu \mathrm{rad_{rms}}$ in the rotation angle. 

\begin{figure}[htbp]
\centerline{\includegraphics[width=10cm]{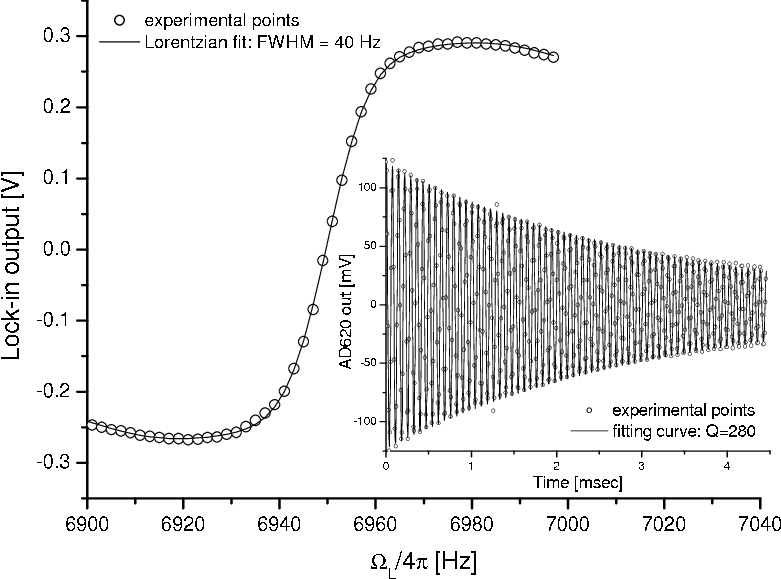}}
\caption{Resonance in the rotation-angle amplitude registered at the
  output of the lock-in amplifier. The output of the instrumentation
  amplifier is fed into the lock-in input, whose reference signal is
  extracted from a pulse generator which modulates the pump
  laser frequency. The resonance linewidth is consistent
  with the quality factor inferred from the exponential decay 
  of the amplitude after a single pump pulse (in the inset). 
\label{biancalana_f4}} 
\end{figure}

A PC controlled stepped ramp of $\nu$ around $\nu_L$ (or $\nu_L/2$, $\nu_L/3$, ...)
in the open-loop configuration shows the resonance in the rotation-angle
amplitude (see Fig.\ref{biancalana_f4}, where the lock-in is referenced
to the second harmonic of $\nu$ while $\nu$ is scanned around $\nu_L/2$).
Optimal magnetometer performance is obtained at a pump beam power of $3~\mu$W and a
detecting laser power of $5~\mu$W at the output of the cells. 
The pump beam absorption is of the order of 50\%  and 
about 20\% for the probe laser at the lock point. In this condition the oscillating
single photodiode current is $I_{p-p}^{ac}=13~$nA at a dc current of
$I^{dc}=0.8~\mu$A. Thus the contrast of the resonance is about
$3\%$ for each polarimeter arm and the width is of the order
of 40~Hz in the second harmonic detection. This width is consistent 
with the quality of the oscillator (Q of 280, as seen from the inset) 
verified under identical probe conditions, with single pump pulse excitation.
Increasing the probe laser power by a factor of two
steepens the resonance slope by 10\% but light shot noise increases at
the same time. Lower probe laser power gradually decreases the slope,
for example at $2.5~\mu$W the slope is 30\% less.

\section{Noise and limits}
The self-oscillating mode is obtained by closing the loop so that the
pump laser frequency modulation is driven by a pulse triggered by the output of 
the scaler (generally we use a divider by two). The overall phase shift is
compensated by adjusting the delay of the pulses. In this regime the
Larmor precession signal is not free from the magnetic noise in the
Laboratory. Low-frequency drifts of the center-frequency readings
can be seen (in the range of up to a few 10~Hz) due to variations 
in the earth's magnetic field, ionosphere activity, human activity
etc. The power-net contribution at 50~Hz and its harmonics represent
the major noise contribution, ranging from a few 100~pT up to a few nT. 

In our arrangement complete compensation of the horizontal components of
the dc magnetic field is performed using two orthogonal pairs of
Helmholtz coils. A third pair is used to reduce 
the {\it z} component down to about 4$~\mu \mathrm T$. Accurate of
compensation of the  $\partial B_z/ \partial y$ gradient component
(the one most responsible for line broadening) is  performed by means of
a quadrupole field generated by a pair of magnetic dipoles symmetrically placed
about 1~m away from the sensor. A second  pair of dipoles 
compensates $\partial B_z/ \partial x$, {\it i.e.} equalizes the field
in the two cells. A detailed description of our dc magnetic field and
gradient compensation system is given in \cite{Bel_cardio07}.  

As stressed above, the magnetometer works in an unshielded environment 
and can work in single or differential regimes. 
In single arm operation the output of the instrumentation 
amplifier can be monitored using a frequency counter, thus providing an
absolute measure of the magnetic field.

Small relative magnetic field variations are registered in 
differential mode by forcing the second arm of the magnetometer with the
Larmor frequency of the main arm. This allows for automatic registration of
magnetic field variations with a linear dynamic range that is limited
by the resonance linewidth. The fact that the two arms are only 7~cm
away from each other makes it possible to reject a large part of the 
magnetic noise registered  as common-mode variations. In this regime 
the instrumentation amplifier output of the secondary arm is
demodulated with respect  to the Larmor precession frequency of the
main arm by means of a lock-in amplifier.

\begin{figure}[htbp]
\centerline{\includegraphics[width=10cm]{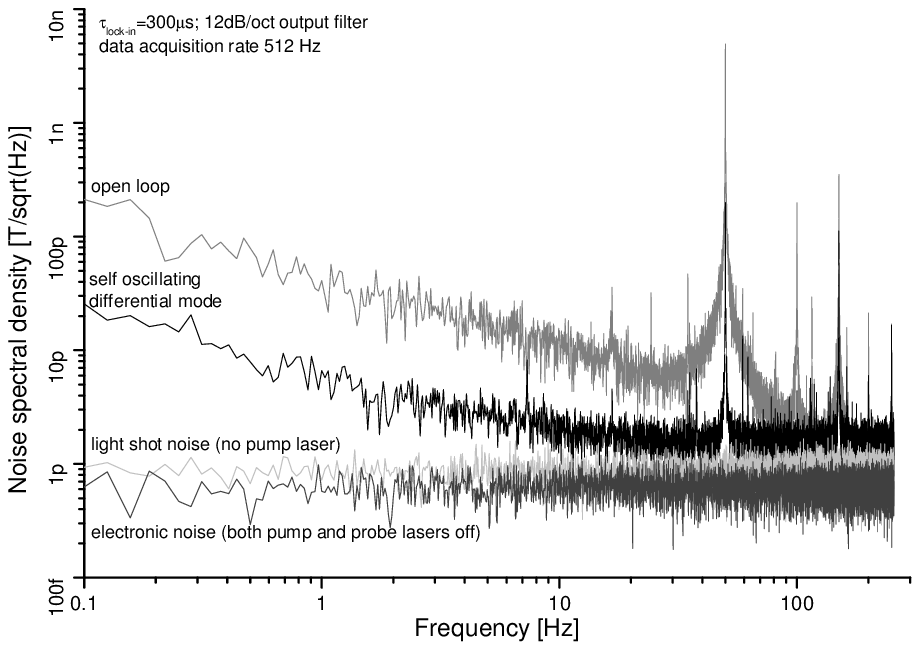}}
\caption {Noise pattern in the open loop, self-oscillating
  differential arrangement, with no pump laser, and the electronic
  noise.\label{biancalana_f5}} 
\end{figure}

The major noise contributions calculated for each detector are as follow: 
the light shot noise
($\sqrt{\mathrm{2qI_{dc}}}=0.5~{\mathrm{pA}}/\sqrt{\mathrm {Hz}}$)  
of \pthz{0.62}, the Johnson noise
($\sqrt{\mathrm{4RTk_B}}=33~{\mathrm{nV}}/\sqrt{\mathrm {Hz}}$) 
estimated to be \pthz{0.6} and the noise of the amplifiers 
($0.4~{\mathrm{nA}}/\sqrt{\mathrm {Hz}}$,
$4~{\mathrm{nV}}/\sqrt{\mathrm {Hz}}$) of \pthz{0.5}. The overall
noise contribution from the four detectors amounts to \pthz{2} which is
consistent with the  measured residual differential noise spectrum
shown in Fig.\ref{biancalana_f5}.  

\section{Registration of water bulk magnetization}
\begin{figure}[htbp]
\centerline{\includegraphics[width=10cm]{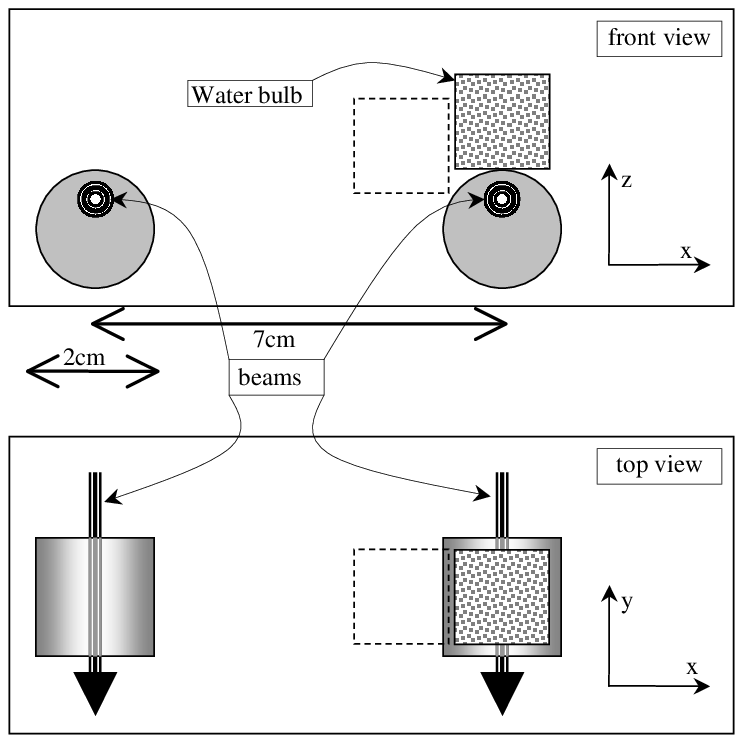}}
\caption{Front and top views of the arrangement of the dual sensor and 
of the water bulb. The grey cylinders represent the Cs cells. Bulk water
magnetization is measured when the bulb (dotted cube) is placed above the cell. 
The dashed cube represents the position of the bulb for detection of
free induction decay.\label{biancalana_f6}} 
\end{figure}

The differential magnetometer is used to detect the bulk
magnetization of tap water in the setup represented in Fig.\ref{biancalana_f6}. 
The water flows through a rectangular cross-section tube ($6\times
8$~mm$^2$ inner size)  in a prepolarizing field. The prepolarizing
field is generated by a set of Nd permanent magnets (1~inch cubes)
aligned so as to produce a pair of 70~cm long magnets. The
prepolarizing magnetic field strength is estimated to be 0.7~T at the
center of the assembly (at a distance of 3.5~mm from the surface of
the cubes). The assembly is inserted into an iron cylinder in order to
reduce the external magnetic field and to increase the internal
one. The water volume inside the magnet is about 30~ml and the water
flow is around 5~ml/sec. Thus, the water spends at least 6~sec in the
prepolarizing magnetic field. At the output of the assembly, water
flows through a capillary tube (internal diameter of 1.8~mm) of 1~m in
length, which brings it to the registration volume in 0.5~sec. The
water is accumulated in a measurement bulb of 5~ml (1~sec complete
refreshment time). Water passes through the bulb in a serpentine
path in order to decrease its mixing in the detection volume  and
therefore guarantee an efficient refreshment. The center of the bulb
is  situated at around 1.7~cm from the center of the probe beam.

The water flow is modulated using a peristaltic pump which is periodically
switched on and off. A typical measurement consists of a set of on-off cycles, 
which are necessary to average out slow drifts, and the high frequency
residual noise. In each cycle, the pump-on duration is 2~sec: this
time is necessary to reach a stable flow regime (0.5~sec), transport the water
to the detection region (0.5~sec), and complete refreshment of the bulb
(1~sec). The pump-off duration is 5~sec, which is necessary to 
let the longitudinal magnetization {\bf M} reach complete
relaxation. The  {\bf M} decay in time is detected by the magnetometer in this
interval. The output of the lock-in amplifier is stored together with
the pump status  in time for further data averaging. 

\begin{figure}[htbp]
\centerline{\includegraphics[width=10cm]{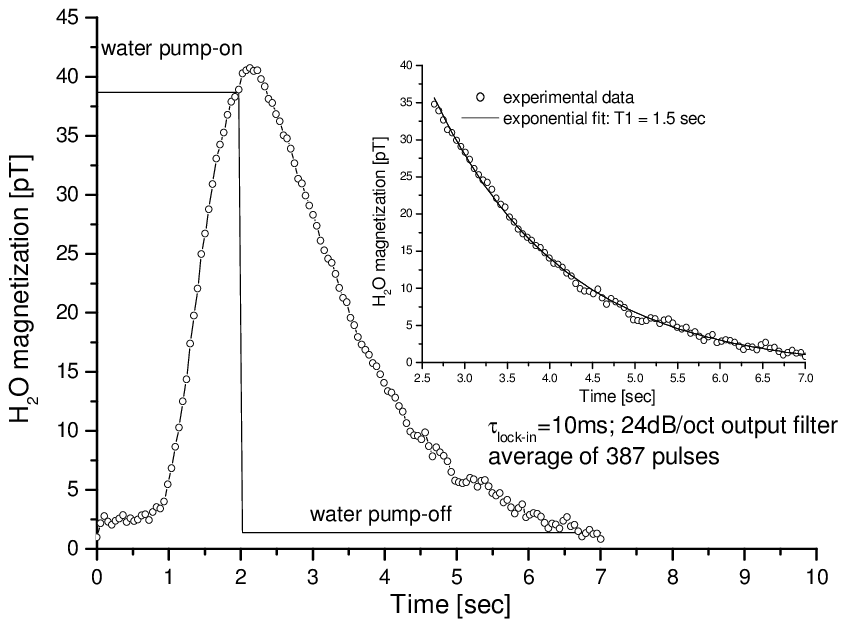}}
\caption{Magnetic field variation in {\it z} direction due to water
magnetization. The signal shown is the result of an average of 387
pulses.\label{biancalana_f7}} 
\end{figure}

The self-oscillating operation makes the magnetometer
insensitive to the slow drifts of the magnetic field (in a range of
several nT), thus making very long recording times possible. 
The stability of the system allows for autonomous recording times
as long as several hours (actually they were usually limited by the
duration of the working day). The average signal of {\bf M} 
of 387 water pump cycles is shown in Fig.\ref{biancalana_f7}. The
estimated $T_1$ relaxation is of the order of 1.5~sec, as shown in the inset.

\begin{figure}[htbp]
\centerline{\includegraphics[width=10cm]{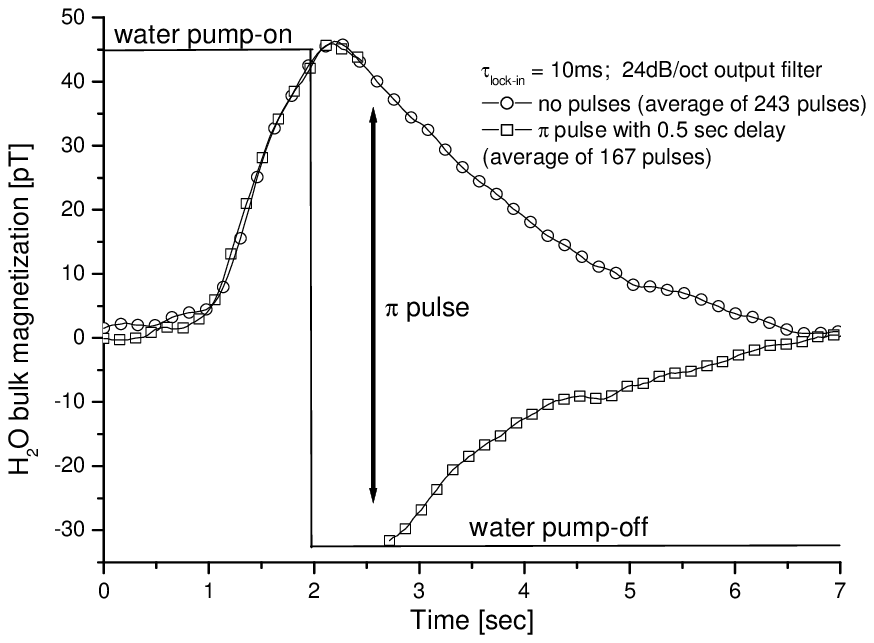}}
\caption{Magnetic field due to water flowing in the detection region
  (circles) and following a $\pi$ pulse (squares). The data are
  recorded with a time constant of 10~ms and a 24~dB/oct output
  filter. The $\pi$ pulse is applied using {\it x} direction
  compensating coils.\label{biancalana_f8}} 
\end{figure}

Water magnetization decay can also be observed via a variant of phase encoding.
An inversion of the magnetization can be observed by applying 
$\pi$ pulse using the compensating coils ({\it x} or {\it y}
pair). Fig.\ref{biancalana_f8} shows the trace of non-perturbed {\bf
M} increase and decay, together with a trace showing its inversion
when a $\pi$ pulse is applied at the end of the pump-on interval after
0.5~sec delay. The $\pi$ pulse is composed of 8 periods of sine waveform 
whose frequency is automatically determined by reading the current lock-in 
reference frequency (the atomic Larmor frequency of the main arm) and scaling it 
by the ratio between the atomic and nuclear gyromagnetic factors.

\begin{figure}[htbp]
\centerline{\includegraphics[width=10cm]{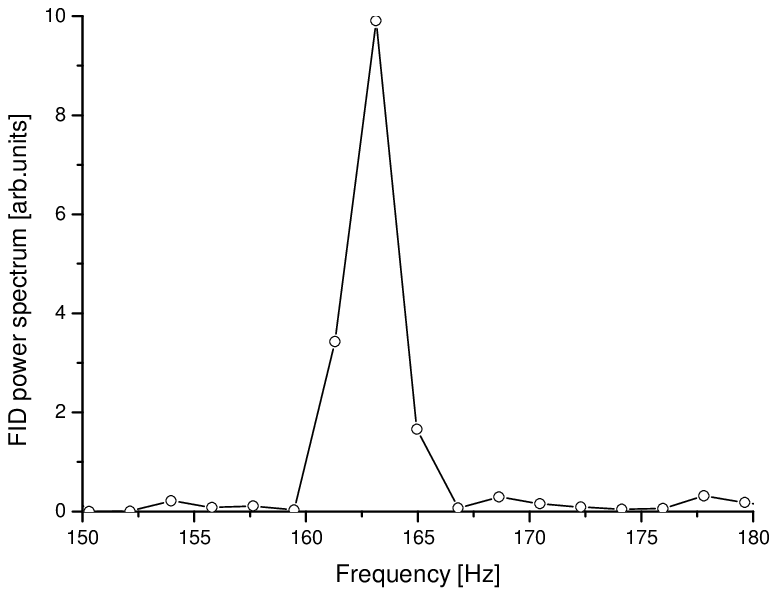}}
\caption{Power spectrum of the magnetic field demodulated
with the atomic Larmor frequency. The peak at 163~Hz corresponds to the 
FID following the $\pi$/2 pulse applied with the {\it x} pair of compensating
coils. The signal is acquired at 30~kS/s and is an average of 43 pulses. 
\label{biancalana_f9}}
\end{figure}

The response of the self-oscillating magnetometer at 
higher frequencies is evaluated by detecting the free induction decay (FID) 
of the nuclear spins. In this case the magnetometer works in single arm 
mode (only using the arm closest to the water bulb).
In order for the FID to contribute with a non-vanishing component
parallel to the bias magnetic field $B_z$ ({\it i.e.} to produce a
signal detectable by a scalar magnetometer), the sample 
is displaced by about 45$^\circ$ with respect to the direction of the
bias field (see Fig.\ref{biancalana_f6}). 
The geometrical constraints (larger distance and angular spread) make the 
signal noticeably weaker than in the dc case (the maximum dc bulk
magnetization  detected is of the order of 30~pT), 
so that accurate signal processing is necessary to extract it. 
In the case of Fig.\ref{biancalana_f9} the procedure consists in
acquiring a set of  traces (direct digitization of the instrumentation
amplifier output) starting at the end of the $\pi$/2 pulse. The signal is
digitized and numerically demodulated with the carrier frequency (at
$\nu_L\approx 13$ kHz), which is evaluated for each single trace. 
Following a 200 ms delay, which is necessary to stop the water
flow completely, the pump-off triggers the $\pi$/2 pulse, at the 
end of which a trace acquisition starts. The data sets are averaged
and analyzed  in order to reduce the amplitude of the noise components
having a random phase with respect to the trigger and to highlight
the nuclear Larmor precession at 163~Hz. Deterministic noise contributions at
50~Hz and its harmonics (up to the 3$^{rd}$ harmonic) are estimated and 
subtracted prior to evaluation of the power spectrum.

\section{Conclusions}
We have demonstrated a two-arms and two-beams self-oscillating 
magnetometer based on nonlinear magneto-optical rotation.
The use of a non-modulated probe beam provides a clear Larmor precession
signal and the use of sub-harmonic excitation prevents  
fake signals in the loop. The optical magnetometer reaches a 
sensitivity of \pthz{2} in unshielded environment.
The differential sensor makes the optical magnetometer robust
and gives it a high dynamic range when working in self-oscillating
regime. It also brings the further advantage of long operating times 
and thus the possibility of significant noise reduction 
when using, for example, an off-line averaging procedure.
The sensor was used for the NMR registration, detecting both the dc 
magnetization and the nuclear spins precession of remotely polarized
hydrogen nuclei in water samples. The low-field operation 
and the corresponding low precession frequency of the nuclei would facilitate
nuclear spin manipulation via non-adiabatic magnetic pulses. It should
be emphasized that such high sensitivity optical magnetometry offers
the possibility of NMR registration using relatively low prepolarizing
magnetic fields.

\section{Acknowledgments}
The authors would like to acknowledge very useful suggestions
about the implementation of the self-oscillating set-up with D.\ Budker,
encouraging discussions with A.\ Cassar\`a about low magnetic
field  NMR detection, and Emma Thorley for her kind collaboration in
improving the manuscript. G.\ B., V.\ B., Y.\ D. and L.\ M.  
would like to acknowledge the financial support of the Monte dei
Paschi Foundation (ref. number 27285).


\begin{thebibliography}{99}
\bibitem{nota}
We use OSD15-5T (Centronic) photodiodes with a sensitivity of 0.45~A/W and 
5~nA dark current. Each photo-current is amplified using an OP484 chip (Analog devices), 
which contains four low-noise operational amplifiers (OAs). One OA is
configured as the transimpedance amplifier, with a feedback impedance made
by a 68~$k\Omega$  resistor in parallel with a 47~pF capacitor. Another OA is
used as a dc voltage amplifier, while the two remaining OAs are used
as cascaded bandpass amplifiers (each one has an ac coupled input and
a  RC parallel in its feedback, with cutoff frequencies set at 10~kHz 
and 40~kHz, respectively).
\end{thebibliography}
\end{document}